\documentclass[final,5p,times]{elsarticle}

\usepackage{epsfig}

\usepackage{amssymb}

\usepackage{lineno}
\usepackage[utf8]{inputenc}

\journal{International Journal of Modern Physics C}

\begin{document}

\begin{frontmatter}

\title{Winning strategies in congested traffic}

\author[ubb]{Ferenc J\'arai-Szab\'o}
\author[ubb]{Zolt\'an N\'eda}

\address[ubb]{Department of Physics, Babes-Bolyai University, str. Kogalniceanu 1, RO-400084 Cluj-Napoca, Romania}

\begin{abstract}
One-directional traffic on two-lanes is modeled in the framework of a spring-block type model. A fraction $q$ of the cars are allowed to change lanes, following simple dynamical rules, while the other cars keep their initial lane. The advance of cars, starting from equivalent positions and following the two driving strategies is studied and compared. As a function of the parameter $q$ the winning probability and the average gain in the advancement for the lane-changing strategy is computed. An interesting phase-transition like behavior is revealed and conclusions are drawn regarding the conditions when the lane changing strategy is the better option for the drivers.
\end{abstract}


\end{frontmatter}
\section{Introduction}

Each time we are stacked in the middle of a traffic jam there is a feeling that the cars in the other lane are advancing better. In such situations the question {\em "To change, or not to change the lane we are advancing?"} naturally turns up in our mind. The right answer to this question is not  straightforward, however.

In order to attempt an answer to this dilemma two possible advancing strategies are analyzed and their efficiency is compared by computer simulations. Our approach is based on a spring-block chain model, which was successful in describing some known aspects of the single-lane highway traffic\cite{Jarai2011,Jarai2012-PhysA}. Here, it has to be noted that for explaining the variety of complex non-linear phenomena present in agglomerated traffic systems many theoretical models have been developed\cite{Chowdhury2000,Helbing2001,Nagatani2002,Kerner2004,Maerivoet2005,Mahnke2005,Darbha2008}. With this in mind, it has to be clarified that our aim is not to give a better approach. In this work, our main goal is to report, based on a simplified null-model, some new aspects of highway traffic and to propose it for further analysis. Apart of this, the present work has also the aim to emphasize the interdisciplinary character of the simple spring-block model family\cite{Neda2011}.

The single-lane spring-block traffic model has been introduced recently\cite{Jarai2012-PhysA}. The blocks model the cars in a lane and springs that acts unidirectionally from the car ahead to the car in the back, model the distance keeping interaction between cars. Using these simple elements a spring-block chain is built. The first car (block) is dragged with constant velocity $v_0$, inducing the movement of the other cars in the chain. With these elements the spring-block chain tends to be equivalent to the well known car-following models\cite{Helbing2001} with instantaneous reaction of drivers. 

It was recently shown\cite{Wang2011}, that the characteristics of individual drivers is also influencing the stability of traffic flow. In this sense, the spring-block models have another important ingredient. This is the friction that acts on blocks and opposing their free sliding. In analogy with classical mechanics the movement of blocks is opposed by a static or a kinetic friction force ($F_S$ and $F_K$, respectively). Extending even more the analogy with classical mechanics, the ratio of these two forces $f = \frac{F_K}{F_S}$ is kept constant. The difference between $F_S$ and $F_K$ models the so called slow-to-start rule first proposed by Barlovic et al.\cite{Barlovic1998}. In the model used for the single-lane highway traffic the values of the static friction forces (and implicitly the values of the kinetic friction forces as well) are generated randomly for any new position of each car, assuming a normal distribution characterized by mean value $\langle F_S\rangle$ and a standard deviation $\sigma$. In this sense, the friction itself models the differences, imperfections and unpredictable reactions of the drivers. This introduces a characteristic disorder in the model, with a major influence on the observed collective dynamics. 

Using of the above sketched spring-block traffic model, in our previous work\cite{Jarai2012-PhysA} the single-lane highway traffic was analyzed. Our studies concluded that in the parameter space defined by the characteristics of the friction forces and drag velocity, $v_0$ two distinct type of dynamics are distinguishable. On one hand, there is a continuous or free flow regime characterized mostly by the continuous motion of the blocks. On the other hand, there is also a congested flow phase characterized by the spontaneous emergence of the so called ''phantom traffic jams''\cite{Schonhof2007,Nakayama2009}. 

It has to be mentioned however that contrary to most of the studies in the field of highway traffic, in our previous study\cite{Jarai2012-PhysA} we have focused on a measure characteristic for one car in the row: the distribution of time intervals during which the car is not moving (stop-time). The two distinct phases have been identified by an order parameter $r$ which is calculated as the ratio of the stop-times standard deviation and the mean stop-time of the selected car. A second-order phase transition separates the two phases\cite{Jarai2012-PhysA}, and it was shown that for each  value of $f$ a critical $\langle F_S \rangle_c$ and $\sigma_c$ may be defined. In the continuous flow regime the order parameter has values around $r = 0.5$ and in the congested flow phase it becomes grater than $r=1$. For a given drag velocity, $v_0$, the transition is realized sharply in a narrow region of the parameter space defined by $\langle F_S \rangle_c$ and $\sigma_c$.

Here, we extend the above discussed spring-block approach to the two-lane traffic situations. The extended model is analyzed and the obtained results are compared to the previous single-lane traffic simulation results. Then, the lane-changing and lane-keeping strategies are analyzed and compared to each other. Based on this comparison the advantages and disadvantages of both strategies are highlighted.

\section{The two-lane spring-block traffic model}

\begin{figure}
\centerline{\psfig{file=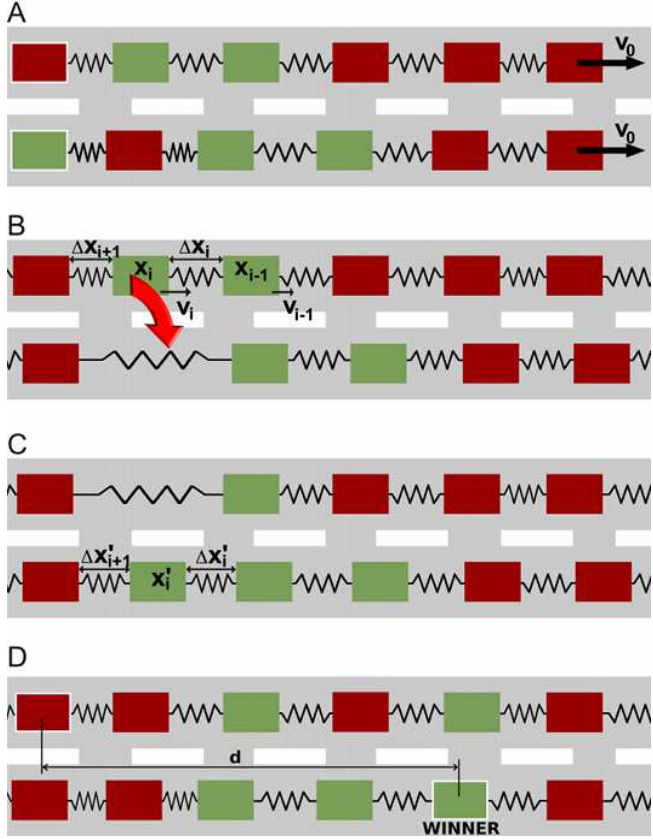,width=8.5cm}}
\caption{Sketch of the lane-changing dynamics in the two-lane spring-block model.\label{fig2}}
\end{figure}

The extension of the single-lane spring-block traffic model for the case of a unidirectional two-lane traffic is straightforward. Two identical spring-block chains are placed close to  each other as sketched in Fig. \ref{fig2}A. The first car of each chain is dragged simultaneously with constant velocity $v_0$. The rest of the blocks are moving according to the rules of the single-lane model\cite{Jarai2012-PhysA}. It has to be noted here that we are interested in such conditions where the two lanes are equivalent in  their dynamics. Contrarily to real highways, there is no rule that states that one of the lanes is used for advancing purposes only. Anyhow, the advancing rules are usually not applicable in case of dense, congested traffic flows, where both lanes are fully occupied. In such view our model becomes applicable in real congested traffic situations. 

In congested traffic two simple driving strategies may be defined. On one hand, there are drivers that follow a lane-changing strategy: they will change the lane whenever they are stacked in the traffic and consider that there is a better advancing possibility in other lane. The blocks corresponding to those cars are drawn with light grey tone in Fig. \ref{fig2} and their strategy is labeled by ST1. On the other hand, there are drivers that follow a lane-keeping strategy. The cars acting with this strategy will never change the lane even if there are stacked in a traffic-jam. Such blocks are drawn with dark grey tones on the Fig. \ref{fig2} and their strategy is labeled as ST2. At the beginning of the simulation one of these two strategies is assigned to each car. The fraction $q$ of cars that are acting with the ST1 strategy is the main parameter of the two-lane model. Its effects will be discussed later. 

Beside the car-following dynamics, driving on a multilane road assumes also a lane-changing process. In the literature, the most general method to include these processes is to introduce lane-changing rules\cite{Nagel1998,Kukida2011} on any car-following model, like for example the cellular automata model\cite{Nagel1992}. The same receipt will be followed in case of the spring-block model, as well. For the sake of simplicity we will proceed with some basic and realistic lane-changing rules that ensures the ``interaction'' of the two spring-block chains. As in the case of many other multi-lane studies\cite{Lv2011}, the rules take into account the headway difference, velocity difference and the safety distance. The defined rules are applied for cars following strategy ST1 only, and they state that in general a driver has reason to change lanes for better driving conditions. Usually, a driver will try to change lanes when there is a car in his/her front that is advancing slower than he/she. The situation is sketched in Fig. \ref{fig2}B. In terms of our spring-block model the rules can be stated as follows:

(1) A block $i$ following strategy ST1 is selected for lane changing, if the distance $\Delta x_i$ to the previous block in the row is smaller than the average distance between cars $d_{avg}$. The distance $d_{avg}$ is measured and fixed at the beginning of each simulation, and it represents the average distance between cars in case of a single spring-block chain that advances under the same conditions (having the same model parameters). 

(2) Also, in order to be eligible for lane changing, the velocity $v_i$ of the car has to be greater or equal to the velocity $v_{i-1}$ of the car ahead. 

(3) The cars eligible for lane-changing  will execute the lane-changing maneuver only if there is enough space next to it on the other lane for the entering safely (see Fig. \ref{fig2}C). This safety criteria states that the both of the corresponding distances $\Delta x'_{j}$ and $\Delta x'_{j+1}$  between cars in the new lane, have to be greater than the minimum following distance $d_{min}$ already defined in the context of the single-lane spring-block model\cite{Jarai2012-PhysA}. 

(4) Moreover, we impose that the distance $\Delta x'_j$ has to be large enough in order to ensure an advance of the newly entered car in the next simulation step. This condition is met if the net spring force acting on the newly entered car will be greater than the friction force in it's new position.

\begin{figure}
\centerline{\psfig{file=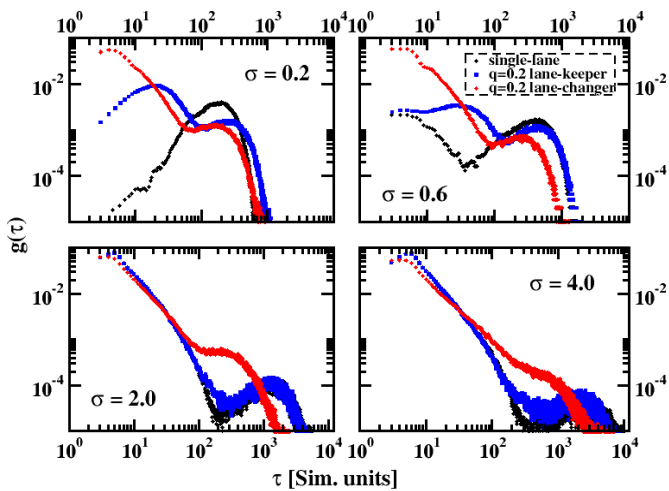,width=9cm}}
\caption{Stop-time distributions of the two-lane model in comparison to the one obtained in the single-lane version.\label{fig3}}
\end{figure}

Apart of using these additional lane-changing rules the dynamics of the single-lane spring-block model\cite{Jarai2012-PhysA} is followed. In order to get a first impression about the dynamics of the blocks in the obtained two-lane model, in Fig. \ref{fig3} the stop-time distribution $g_{2l}(\tau)$ of an ST1 and ST2 block, both positioned at the end of the initial queue, is compared. Comparison is also made with the typical stop-time distribution $g_{1l}(\tau)$ of the single-lane model\cite{Jarai2012-PhysA}. In this  initial study only a fixed portion ($q=0.2$) of cars with ST1 strategy are considered. The other model parameters $\langle F_s \rangle = 4$ and $f=0.8$ are fixed and $sigma$ is varied. 

Stop-time distribution functions are constructed from $100000$ simulated stop-times and are plotted for different disorder levels, $\sigma$, on the panels of Fig. \ref{fig3}. As it is immediately observable, at low disorder level (top left panel) both the ST1 and ST2 distributions of the two-lane model are dominated by shorter stop-times relative to the single-lane case. This confirms that in case of free-flow (or jam-free) traffic the presence of a second lane improves the advance of the cars. The difference in the stop-time distribution for the ST1 and ST2 strategies is important only in the very short stop-time limit. 

However, at higher disorder values, where the traffic becomes jammed, there is a more significant difference between the stop-time distributions of the ST1 and ST2 cars. The distribution function for cars with ST2 strategy looks exactly the same as the distribution function constructed in case of a single-lane traffic model. In contrast, the maximum from the long stop-time regime in case of ST1 cars is shifted into forward direction. This suggests a better advance of them because they perform in general shorter stops than the ST2  cars. These results gives us a first qualitative impression about the differences between the strategies. In addition, the results also shows us that due to the complexity of the system, in the jamming phase further investigations are needed, and the influence of the $q$ parameter has to be also studied. 

\section{Comparison of advancing strategies}

In the framework of the two-lane spring-block model we are interested in comparing the efficiency of ST1 and ST2, namely the lane-changing and the lane-keeping strategies.  As it was suggested in the previous section, in case of free-flow conditions  one expects no significant differences between the stop-time distributions corresponding to these strategies. Accordingly, in the following our attention will be focused to the jam\-med traffic conditions only. 

Let us fix for this study parameter values that leads to jam\-med traffic conditions in the single-lane model\cite{Jarai2012-PhysA}: $\langle F_s \rangle = 9.6$, $\sigma = 2.6$ and $f=0.8$. For selecting these parameter values our requirement was to be far from the critical parameter values that defined the phase transition in the single-lane system. In other words this means to be deeply inside the jammed traffic phase. By fixing these parameters, the only free parameter remains the ratio $q$ of the ST1 strategies in the row. In order to test and compare the efficiency of the presumed strategies computer simulations on the two-lane model have been performed. In these studies the last car of the first row is set to follow the ST1 strategy while the last car of the other lane is set to advance respecting the ST2 strategy. The position of the selected cars are compared at the end of each simulation. At the beginning of each simulation the advancing strategy of the other cars in the queue is set to ST1 with probability $q$ and to ST2 with probability $1-q$. After initializing the system, the first car of both rows is dragged in a parallel manner through a distance of $D=5000$ simulation units which corresponds to a real distance of approximately $20$ km. The details of this conversion are explained in our previous work\cite{Jarai2012-PhysA}. Finally, after completing this distance, the positions of the selected cars are compared to each other, as shown in the Fig. \ref{fig2}D. The car that is closer to the beginning of the row is declared as winner. The simulations are repeated and from a statistics of $1000$ simulations the winning probability $w$ of the lane-changing strategy is determined.

\begin{figure}
\centerline{\psfig{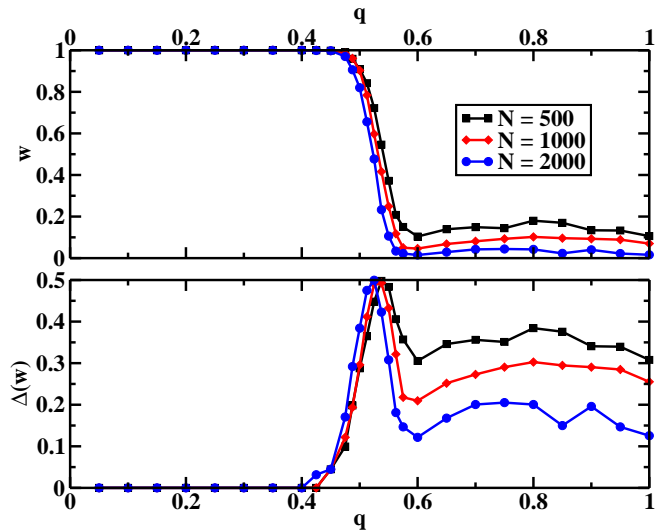}}
\caption{Winning probability, $w$, of the ST1 strategy as a function of the ratio $q$ of cars following this strategy. The figure from the bottom shows the fluctuation of the order-parameter $w$ as a function of $q$.\label{fig4}}
\end{figure}

Simulation results for three different lengths of the chain ($N = 500; 1000$ and $2000$ blocks on each lane) are plotted on the top panel of Fig. \ref{fig4}. For $q<0.5$ the winning probability $w$ of the car following the ST1 strategy  is $w=1$. On the other hand, the same winning probability for $q > 0.5$ sharply decreases to small values. The transition regime is narrow and the transition becomes sharper as the number of blocks, $N$, is increased. Moreover, in the $q > 0.5$ interval, the winning probability clearly decreases with the size of the system. On the bottom panel of Fig. \ref{fig4} the standard deviation of the winning probabilities is represented. As expected, the curves show a clear peak at the critical ratio $q_c$. The position of the maximum is slowly shifted toward $q_c \approx 0.5$ as the system size is increased. These results suggests that in the thermodynamic limit the chosen order parameter will jump in a non-continuous manner at the critical point $q_c \approx 0.5$, suggesting a  first-order phase transition.

\begin{figure}
\centerline{\psfig{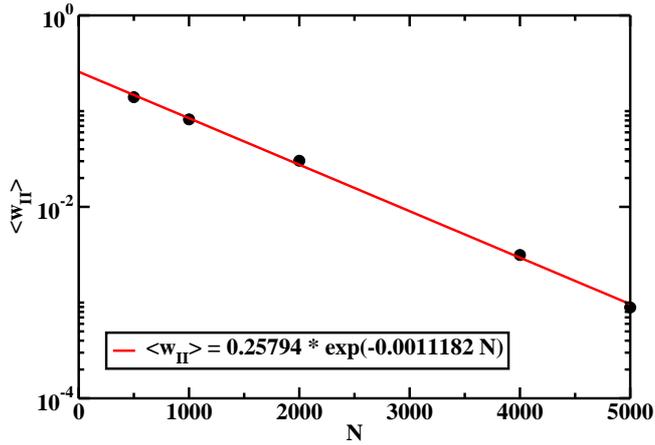}}
\caption{Finite size effects for the winning probability in the $q>0.5$ regime.\label{fig5}}
\end{figure}

For different system sizes (chains formed by different number of cars $N$) the simulations suggest that in the first phase ($q<0.5$) the winning probability is independent of the system size. However, in the second phase ($q>0.5$) , $w$ is monotonically decreasing with the size of the system. Finite-size effects in this region are investigated in detail and the results are shown in the Fig. \ref{fig5}. The average winning probability of the ST1 ($\langle w_{II}\rangle$) is calculated as a function of the system size $N$. The results presented on a log-normal plot suggest an exponentially decaying trend, indicating that for infinite system sizes $\langle w_{II}\rangle \rightarrow 0$.  This result confirms again the presence of a first-order phase transition.

\begin{figure}
\centerline{\psfig{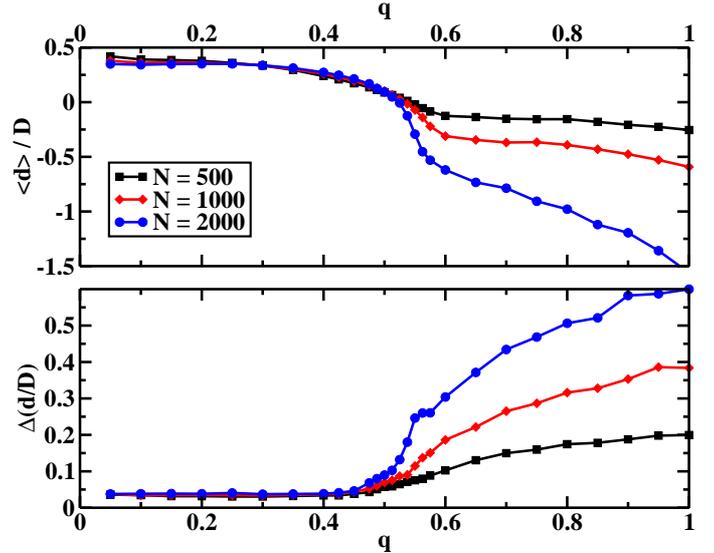}}
\caption{Average gain in the normalized distance ($\langle d \rangle / D$) between the cars following the ST1 and ST2 strategies,
 as a function of the fraction $q$ of cars using the ST1 strategy. Results for different system sizes.\label{fig6}}
\end{figure}

The results may be confirmed by using another characteristic measure, namely the average winning distance of the ST1 strategy. In order to get a value which is independent on the dragging distance $D$, the plotted distances are scaled relative to this value. The results are presented in the Fig. \ref{fig6}. In the top panel of the Figure the scaled average winning distance $\frac{\langle d \rangle}{D}$ is plotted as a function of the fraction $q$. The plot shows that for $q < 0.5$ the distance is positive meaning that the ST1 is the clear winner. On the contrary, in case of $q>0.5$ the distance quickly becomes negative indicating that the other strategy becomes the better one.  The transition is visible in the standard deviations, too. Results in such sense are presented in the bottom panel of Fig. \ref{fig6}.  

\section{Conclusions}
A two-lane traffic model has been built using two identical spring-block chains and simple lane-changing rules. The behavior of the resulted model system was studied in the congested traffic regime. The efficiency  of two driving strategies, one with lane-changing dynamics (ST1) and another with a lane-keeping strategy (ST2) was compared. A clear conclusion emerges from our results, the winning strategy is the one adopted by the less number of cars. If the majority of drivers use the ST2 strategy then the ST1 strategy is the winner and inversely if the majority of blocks use ST1, then the ST2 strategy proves to assure a better advance under congested traffic situations. The model is a first approach for a quite complex phenomenon, and in reality the problem proves to be more complicated. Drivers are not stuck to one of these strategies, and their driving style continuously changes, adapting to what they have previously experienced. The model can be made more complex incorporating these elements, but this is beyond the scope of the present work. The results of the present model can be interpreted however in a statistical sense. This would suggest that even if the strategies of drivers are changing, the important parameter is just the ratio of the two driving styles. The best strategy would be than to select the one what we see less often in the congested traffic. 

In conclusion, the aim of the present work was to investigate by a simple null-model which is the best driving strategy in a congested traffic situation: to change lane whenever is possible, or to get stuck in one lane and continue there. A simple spring-block model approach was considered, which suggests an interesting first-order phase-transition, and thus a simple answer for the problem. The extension of the original spring-block traffic model to the two-lane traffic situation proves again the wide interdisciplinary modeling potential of the classical spring-block type systems.

\section*{Acknowledgments}
This work was supported by the Romanian National Research Council (CNCS-UEFISCSU), project number PN II-RU PD\_404/2010.


\end{document}